# Photosensitive Gaseous Detectors for Cryogenic Temperature Applications

L. Periale[1], V. Peskov[2], C. Iacobaeus[3], B. Lund-Jensen[4], P. Picchi[1], F.Pietropaolo1, Rodionov[5]

[1]CERN, Geneva, Switzerland
[2]Pole University Leonardo de Vinci, Paris, France
[3]Karolinska Institute, Stockholm, Sweden
[4]Royal Institute of Technology, Stockholm, Sweden
[5]Reagent Research Center, Moscow, Russia

**Abstract**

There are several proposals and projects today for building LXe Time Projection Chambers (TPCs) for dark matter search. An important element of these TPCs are the photomultipliers operating either inside LXe or in vapors above the liquid.

We have recently demonstrated that photosensitive gaseous detectors (wire type and hole-type) can operate perfectly well until temperatures of $LN_2$. In this paper results of systematic studies of operation of the photosensitive version of these detectors (combined with reflective or semi-transparent CsI photocathodes) in the temperature interval of 300-150 K are presented. In particular, it was demonstrated that both sealed and flushed by a gas detectors could operate at a quite stable fashion in a year/time scale. Obtained results, in particular the long-term stability of photosensitive gaseous detectors, strongly indicate that they can be cheap and simple alternatives to photomultipliers or avalanche solid-state detectors in LXe TPC applications.

## 1. Introduction

The origin of dark matter is one of the fundamental problems of modern physics. There are theoretical predictions that dark matter consists of Weakly Interacting Massive Particles (WIMPs). Several WIMP detector concepts were developed and tested, see for example [1-9]). One of the most promising detectors could be the one which is based on a LXe TPC which has potential for a unique rejection power [7-9]. An important element of this TPC is the large area of the photmultipliers (PMs) operating inside the liquid or placed in vapors above the liquid. The use of these detectors considerably increases the overall cost of the LXe TPCs and may also bring some additional radioactive background.

There have been several efforts to replace PMs by avalanche solid state detectors [10] however, the cost of these new devices is still high.

In our recent pilot study works we have experimentally demonstrated that gaseous detectors: single wire counters or hole-type detectors could operate until $LN_2$ [11,12].

The aim of this work is to perform systematic studies of gaseous detectors combined with reflective or semitransparent CsI photocathodes in the temperature interval of 300-165 K in order to check if their characteristics match the requirements for the LXe TPC applications. A special focus in these studies was put on long-term stability, which is an essential point in practical applications. The other important issue to address was the operation of the sealed detectors and the operation of the semitransparent CsI photocathodes which may also improve the light collection that is to be achieved better than with the reflective one.

## 2. Experimental Set Up

Our experimental set up is shown schematically in Figure1. It contains a cryostat (see [12] for more details) inside which a "scintillation chamber" was installed, a gaseous detector and a PM or a UV source .The "scintillation chamber" was a cylindrical vessel (having adiameter of 40 mm) with two $MgF_2$ windows on opposite flanges to each other. The chamber was filled either with an Ar or an Xe gas at a pressure of 1 atm. Inside the "scinillation chamber" one of the following radioactive sources were installed: $^{241}$Am, $^{109}$Cd or $^{55}$Fe. The "scintillation chamber" could be independently cooled by being immersed into the dewar filled with alcohol cooled by $LN_2$. This allowed, if necessary, a LXe layer of a few mm in thicknes to be obtained inside the scintillation chamber, fully covering the radioactive source. In these experiments the high purity gaseous Xe was used for condensation to avoid any attenuation of the scintillation light inside the LXe. To one of the $MgF_2$ windows a gaseous detector was attached and to the opposite $MgF_2$ window a stainless steal tube was connected and flushed, depending on measurements either with Ar or $CH_4$ at a p=1 atm. To the opposite end of this tube a PM (Schlumberger 541F-09-17) or a UV source (a pulsed $H_2$ lamp [13]) was mounted.

Two types of photosensitive gaseous detectors were tested and studied: wire- type and hole-type detectors.

The wire-type detectors were either of a single-wire counter flushed with Ar+$CH_4$ at a pressure of p=1 atm (see Figure 2a) or of a sealed wire counter filled with the same gas at a pressure of p=1 atm. The cylindrical stainless steal cathodes of these detectors (having a diameter of 30 mm) facing the $MgF_2$ window were coated by a 0,4 μm thick CsI layer. In the case of the sealed detector, to ensure a high degree of cleanliness, it (the detector) was heated to 60°C and continuously pumped for 2 weeks at a vacuum of better than $10^{-6}$ Torr before being filled with the gas. Some detectors tested were also done with semitransparent CsI photocathdeodes. In this case the metalised $MgF_2$ window (a Cr layer 5 nm in thickness) was coated by an CsI layer 20 nm in thickness. Semi-transparent CsI photocathodes could be easily damaged by air so their installation inside the detector was done in a glow box flushed by Ar after which the detector was pumped to $10^{-6}$ Torr for a few days and heated to 30°C. Only after this procedure being done was the detector flushed by a mixture of high cleanliness.

The following hole-type detectors were tested: the Gas Electron Multiplier (GEM), Capillary Plates (CPs) and Home-Made CPs (HMCPs). The description of the GEM and the CPs used in our experiment is given in [14, 15] respectively. The HMCPs were made of a gold coated G-10 plate 1mm in thickness in which holes of diameters of 0,5 mm were drilled on a pitch of 1 mm. The cathode of the hole-type detectors facing the scintillation chamber were coated by a 0,4 μm thick CsI layer; their anodes were mechanically and electrically connected to the readout plate, see Figure 2b.

As in the case of the single wire counter, these detectors were either flushed by the gas (Ar+10%$CH_4$ or Xe) or filled by a gas and sealed.

Some tests were also done with semi-transparent CsI photocathodes. Their preparation procedures were similar to the one described above.

The quantum efficiency (QE) measurements of our detectors were performed at the wavelength interval of 165-175 nm by using a scintillation light from Xe (produced by an alpha source $^{241}$Am ) or a pulsed UV light from the $H_2$ lamp (the spectrum of which had a sharp peak at 165 nm). Both of the UV sources were strong enough to produce a signal in our detectors even at gas gains of A=1. The absolute intensity of the light beam was determined by a calibrated CFM-3 counter ( see [11]).

## 3. Results

Figure 3 shows the gains ($A_m$ and $A_f$) and the QE variations with the temperature (T) for the wire–type detectors with reflective CsI photocathodes flushed by Ar+10%CH$_4$ at a p=1 atm. The gain $A_m$ was defined as the maximum achievable gain at which a corona discharge appeared; the gain $A_f$ is the gain at which photon feedback pulses appear, ~10% compared to the main pulse (-see [16] for more details). The "characteristic" voltage $V_c$ on the plot is the voltage at which a gain of $10^4$ was achieved. From the data presented in the figure one can see that $V_c$, $A_m$ and $A_f$ increased with a decrease of the T whereas the measured (practical) QE dropped with the decrease of T. These changes were the result of the gas density increases that came with the cooling (see the discussion). Indeed in the case of the sealed single wire detector the values of $V_c$, $A_m$, $A_f$, the QE did not change with the temperature. The important conclusion one can draw from such data is that wire type detectors combined either with reflective or with semitransparent CsI photocathodes are able to operate stably at LXe temperatures at gains high enough to detect single photoelectrons. These detectors were used by us and by the Berkeley group [17] to detect the scintillation light from the LXe. (see also [18].

As an example Figure 4 shows oscillograms of signals from the wire detector with a reflective CsI photocathode and from the PM. recording simultaneously the scintillation light from LXe. One can clearly see that the signal to noise ratio was much better in the case of the wire detector than in the case of the PM.. More studies on this topic can be found in [17].

Figures 5-7 show the $A_m$, QE and $V_c$ for the bare CPs, GEMs and HMCPs as well as for CPs, GEMs and HMCPs combined with CsI photocathodes. In this figure $A_m$ was not exactly the gain at which the breakdown appeared, but the gain at which first signs of unstable behavior appeared. In Figure 5 the characteristic voltage $V_c$ corresponded to the gain of $10^3$, in Fig. 6 the $V_c$ corresponded to a gain of 500 and in Figure 7 the $V_c$ corresponded to a gain of 65. To make the comparison easier we plotted in Figure 8 the maximum achievable gains for all tested hole-type detectors. On the same plot the QE of the CP and the HMCP are presented as well.

The main conclusions one can make from this data are the following:
1) Bare hole-type detectors operate at gains less than the single wire counters, 2) in Ar+CH$_4$ gas mixtures, CPs are able to operate at higher gains than the GEMs, 3) HMCPs can operate in pure Xe at gains of 300-1000, 4) the maximum achievable gain of the hole-type detectors combined with ScI photocathdes is almost 10 times less than in the case of bare detectors. Thus hole type detectors could be in principle used for the detection of the LXe scintillation light, but several detectors operating in tandem are necessary to achieve gains of $A>10^4$ sufficient for the single photoelectron detection.

The tests of cascaded hole type detectors are in progress for the moment and the first results indicated that they can provide high gains, however, only by the cost of more complicated designs.

Figure 9 shows the results of the long-term stability tests. One can see that in the case of the flushed single wire detector with the reflective CsI photocathode its QE dropped from 25 to 20% during 520 days of continuous operation. In the case of thesemitransparent ones, the QE dropped rather quickly during the first days after which it degradated rather slowly.

A very good stability was achieved with a sealed detector. After pumping and heating, it had 16% of the quantum efficiency. After cooling to room temperature (but remainnig pumped) it becames12% and after being filled with the gas, the QE became 10% after which it remained almost unchanged for 520 days.

In contrast to single wire detectors, hole type detectors were tested for 150-200 days only. One can see that during this period fast degradation was observed in the beginning but was then considerably slowed down.

The general conclusion one can make is that the both types of detectors flushed by a gas exhibit some small degradation. In contrast, the sealed detector was extremely stable.

## 4. Discussion

Several important results were obtained with this work.

The main positive results obtained in this work are that both wire type and hole–type detectors can operate stably at LXe temperatures. However, their behaviour with the temperature is very different. For example, measurements performed in this work revealed several interesting facts. One of them is that the maximum achievable gain of wire type detectors flushed with a gas increases with a the temperature decrease, whereas in the case of bare hole type detectors $A_m$ drops rather strongly with a decreasing the temperature. In the case the hole-type detectors combined with CsI photocathodes gains $A_m$ were almost ten times less and do not change with the T. The explanation for this observation could be as follows. It is known that in the case of the gaseous detectors combined with the CsI photocathodes the breakdown occurs via a so called "slow breakdown mechanism" [16,19]. The condition of the discharge appearance is

$A_m\gamma=1$, where $\gamma$ is a probability for the secondary processes. As it was shown in [20, 21], $\gamma$ depends on several parameters, for example $\gamma$ decreases with a decrease in the $V_c/\rho$ ratio, where $\rho$ is the gas density. As one can conclude from the data presented in Figure 3 that during the cooling of the flushed detector the ratio $V_c/\rho$ decreased, thus $\gamma$ decreased as well. Therefore $A_m$ should increase and this was observed experimentally (see Figure 3). In the case of bare hole–type detector, the breakdown appeared via a fast mechanism [14]. The condition for this type of breakdown is $A_m n_0=Q(\rho, d)$ [12], where $n_0$ - is the number of primary electrons created inside the detector by the external ionisation and Q is a constant which depends on $\rho$ and d is the size of the amplification gap. Usually Q increases with d and drops with $\rho$. Thus when the density increases, the $A_m$ decreases. In the case of the hole-type detectors combined with the CsI photocathodes at low gas densities, slow breakdown dominates, however, with the increase of $\rho$ a mixed slow-fast breakdown may appear as well. This is why one can expect that the $A_m$ may first slightly increase with the temperature drop (as in the case of the wire detector) and then may begin dropping when fast breakdown interferes with the "slow breakdown". The observed behaviour of the $A_m$ was close to predicted.

An interesting question arises as to why the maximum achievable gain is higher than in the case of the hole type detectors. Actually, feedback pulses in the wire detector appear at gains of $\sim 3 \times 10^4$, however, due to the large drifting distance to the anode wire they are well separated in time (a few $\mu$s). In the case of the hole-type detector, the drifting distance for the secondary electrons d is small and secondary avalanches contribute much stronger to the building of the positive ions' space charge, accumulation of which may finally lead to the fast breakdown. By the way, the strength of the electric filed on the hole edges is very high and as it was already metioned $\gamma$ increase with the electric field.

The other important results are that semitransparent photocathodes operate stably at LXe temperatures. Detectors with the semitransparent photocathodes may allow better light collection to be achieved for the detectors operating with the LXe TPC.

The next important result is that sealed gaseous detectors operate extremely stably in a time scale of 1,5 years. This opens the possibility of using them in real LXe TPCs.

Finally, it was demonstrated that HMCPs combined with the CsI photocathodes could operate in pure Xe. However, cascaded detectors should be used to achieve high gains; the tests of such cascade detectors are in progress now and preliminary results are encouraging [22]. However, special tests should be done to demonstrated that cascaded windowless hole-type detectors will be able to operate in turbulent Xe vapors above LXe.

**Conclusions**

Several new results were obtained in this work:
1) For the first time it was demonstrated that <u>sealed photosensitive</u> gaseous detectors (single wire and hole-type) combined with <u>semitransparent CsI</u> photocathodes can operate at LXe temperatures. Semitransparent CsI photocathodes allow much better light collection to be achieved.
2) Single wire detectors combined with semitransparent CsI photocathodes can reach gains sufficient enough to detect single photoelectrons.
3) HMCPs with reflective CsI photocathodes can operate in pure Xe and thus could be used in vapors above the liquid Xe. However, several of such detectors operating in cascade mode are required to reach high gains. Special tests are required to verify that the cascaded detector will be able to operate in turbulent Xe vapors above LXe.
4) For the first time <u>long-term tests</u> (of up to 1,5 years) for photosensitive detectors (sealed and flushed by a gas) were performed.

Obtained results show that photosensitive gaseous detectors (with windows and without) could be cheap and simple alternatives to PMs or avalanche solid-state detectors in LXe TPCs. The other potential advantage could be the possibility of manufacturing them from materials having low levels of radioactivity.

**Figures:**

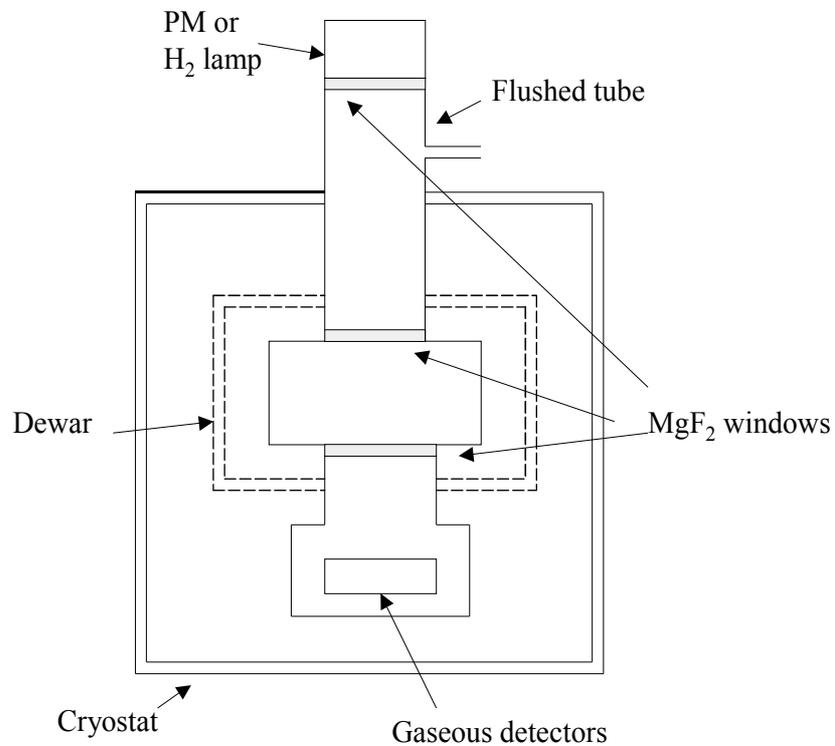

Figure 1. Schematic drawing of the experimental set up

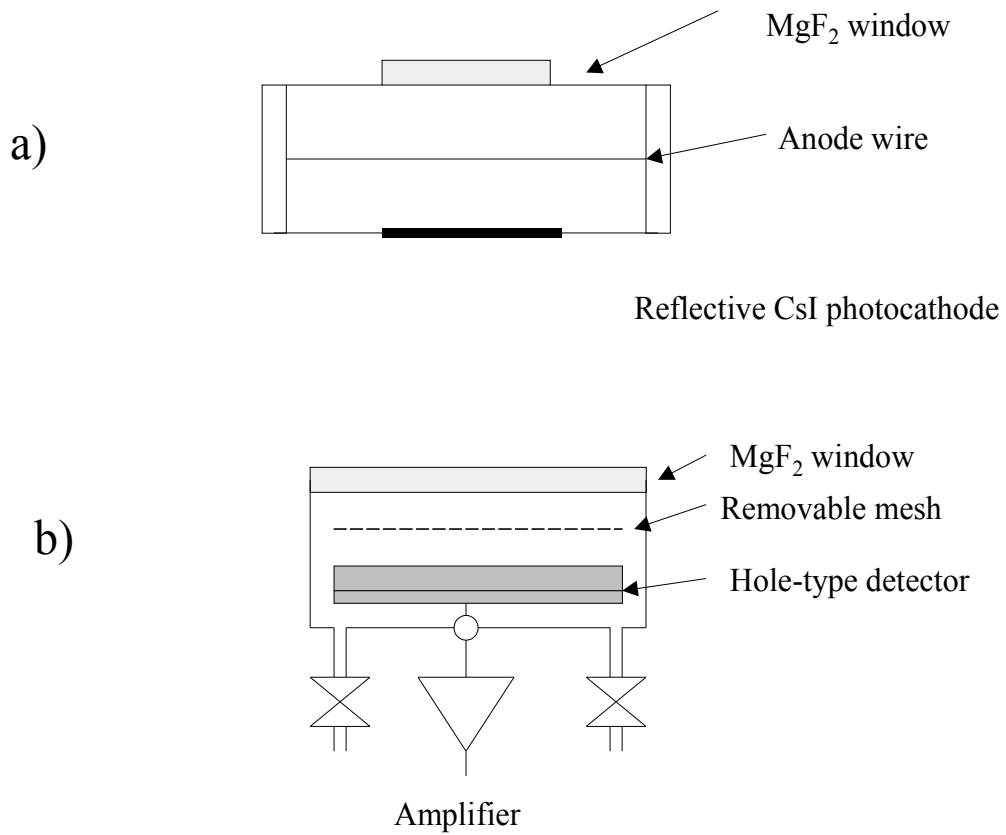

Figure 2. Schematic drawing of the detectors used a) single wire counter b) hole-type detector installed in the gas chamber with the $MgF_2$ window

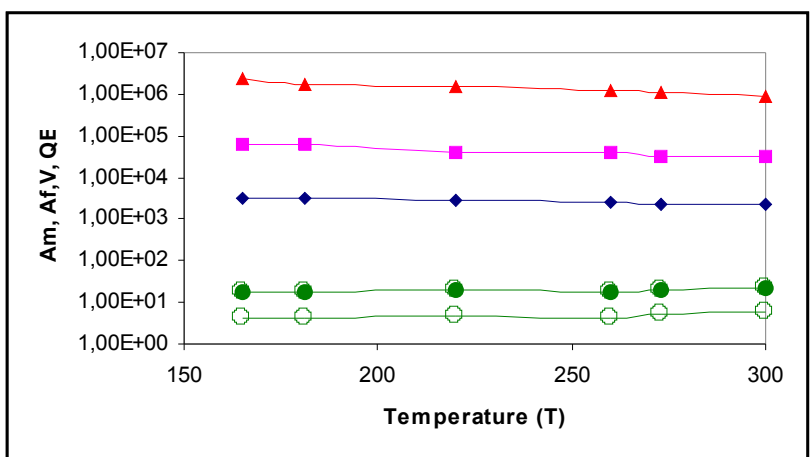

Figure 3. Changes of single wire characteristics with the temperature: red curve-maximum achievable gains, violet curve-gain at which 10% of the feedback pulses appear, blue curve-a characteristic voltage $V_c$ corresponding to the gain of $10^4$, green curve with filled dots-the quantum efficiency of the reflective CsI photocathodes, green curve with transparent dots-semi-transparent CsI photocathodes

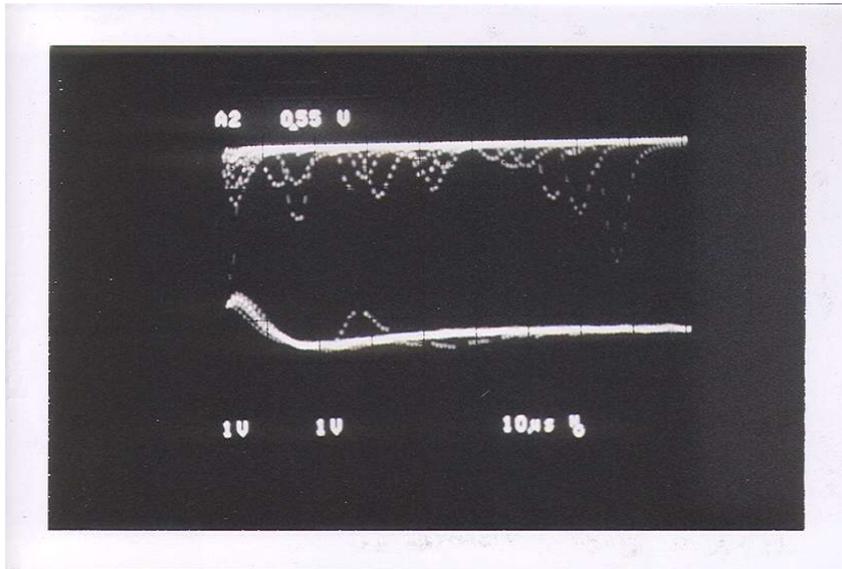

Figure 4. Oscillograms of signals from the PM (upper curve) and the wire-type detector (lower curve) detecting the scintillation light from the liquid Xe

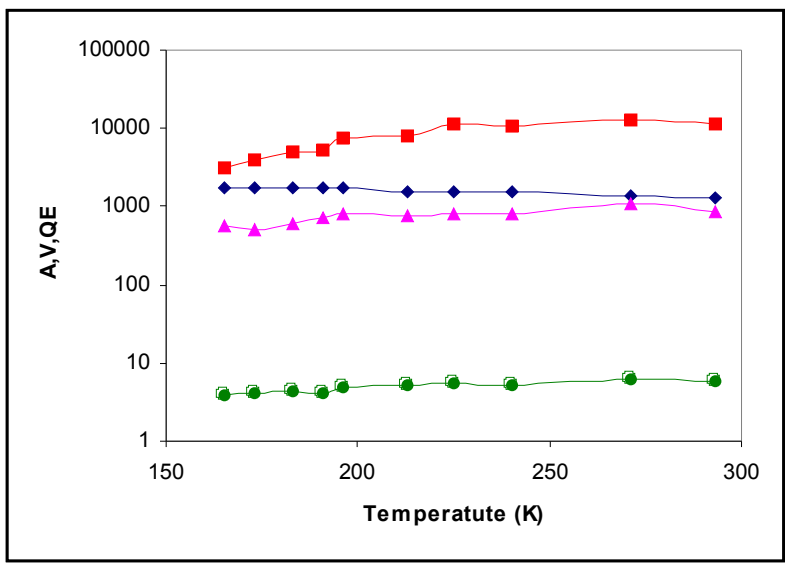

Figure 5. Results obtained with CPs. Red curve-the maximum achievable gain of a bare CP, blue curve-characteristic voltage $V_c$ at a gain of $10^3$, violet curve-gain at which some instabilities appear for the CP coated with CsI, green curve-the quantum efficiency of the reflective CsI photocathode.

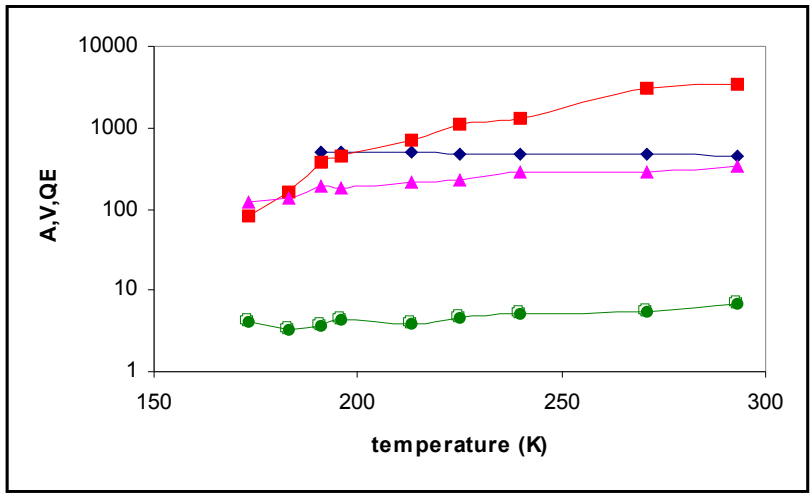

Figure 6. Results obtained with GEMs. Red curve-the maximum achievable gain of a bare GEM, blue curve-characteristic voltage $V_c$ at a gain of 500, violet curve-gain at which some instabilities appear for the GEM coated with CsI, green curve-the quantum efficiency of the reflective CsI photocathode

Figure 7. Red curve-the maximum achievable gain of a bare HMCP, blue curve-characteristic voltage $V_c$ at a gain of 65, violet curve-gain at which some instabilities appear for the HMCP coated with CsI, green curve-the quantum efficiency of the reflective CsI photocathode

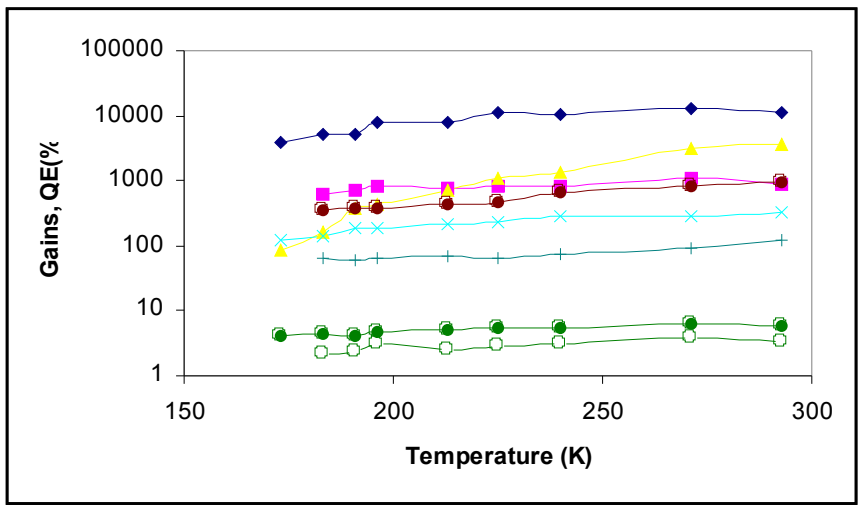

Figure 8. Comparison of the main characteristics of all hole-type detectors: dark blue curve-the maximum achievable gain of a bare CP, violet curve-gain at which instability in operation appears for the CP coated with CsI, yellow curve-maximum achievable gain for a bare GM, light blue curve-gain at which instability in operation appears for the GEM coated with CsI, brown curve-maximum achievable gain for the bare HMCP operating in pure Xe, blue curve-maximum achievable gain for the HMCP coated with CsI, green curve-quantum efficiency of the CP, green curve with filled dots-quantum efficiency of the CP, green curve with clear dots-quantum efficiency of the HMCP in Xe

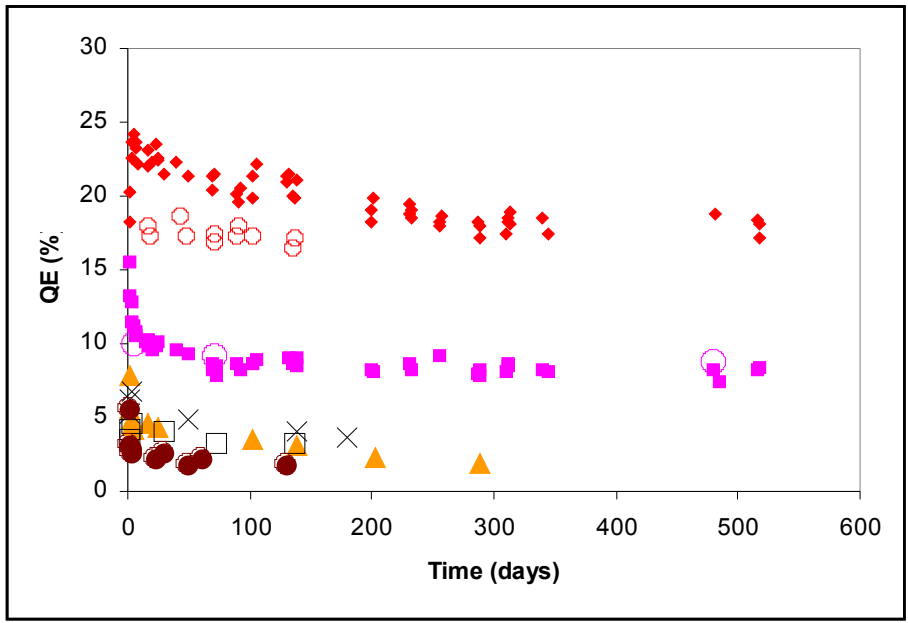

Figure 9. Results of long term stability tests: red rhombus-quantum efficiency of the sealed wire-type detector with reflective CsI photocathodes flushed with with Ar +10% $CH_4$, open red circles-quantum efficiency of flushed wire detectors cooled to liquid Xe temperatures, violet squares-quantum efficiency of the sealed wire type detector at room temperature, violet open circles-quantum efficiency of the same detector cooled to LXe temperatures, yellow triangles-quantum efficiency of the wire type detector combined with semi-transparent CsI photocathodes and flushed with Ar +$CH_4$, black crosses-quantum efficiency of the CP covered by CsI photocathodes, open squares-quantum efficiency of the HMCP in Xe, brown circles-quantum efficiency of the HMCP combined with semi-transparent CsI photocathodes